\documentclass{article}

\usepackage{array,hhline,amssymb,amsthm}

\def\tr{{\rm tr}}
\def\Co{{\rm Const}}

\title{Gravity on the octonion algebra}

\author{V.~Dorofeev\thanks{Friedmann Laboratory For Theoretical Physics,
Department of Mathematics, SPb EF University, Sadovaya 21, 191023
St.Petersburg, Russia, E-mail: dor@vd8186.spb.edu}}

\begin{document}

\maketitle

\begin{abstract}
Gravitational interactions are treated as non-associative part of a
Lagrangian of octonion fields. The Lagrangian is defined in the
charge space as squared curvature with respect to the octonion
fields. The applications of suggested formalism to homogeneous and
isotropic space are studied.
\end{abstract}

\section*{Introduction}

Generalizations of physical theories to octonions were studied in
various aspects (see \cite{Baez} for a review). However the main
problem is that we have to speak about physical interpretations of
new algebraic models having in mind possible inadequacy of the
physical model to its mathematical counterpart, which is primarily
the effect of non-associativity of the algebras in question
\cite{Baez}. Suitable matrix representations of non-associative
algebras, suggested in \cite{Zorn,Daboul} was a reasonable step
forward, but the price for this advancement was a peculiar
multiplication law (\ref{Dab}). As a consequence, in Feinmann's
interpretation of the motion in external field, the state of a
moving quantum particle, even in the absence of charges, is subject
to change caused by the non-associative character of the
interactions. It is reasonable to associate these interactions with
the gravitational field. A holistic picture of gravitation field
arises along these lines, which is the main object of this paper.

\section{Non-associativity and curvature tensor}

The free Dirac equations for spinor fields $\Psi(x)$ and
$\overline{\Psi}(x)$, considered in Minkowskian space $M_4$ with the
metric
\begin{equation}\label{M4}
ds^2=\eta_{ab}dx^a dx^b=dt^2-dx^2-dy^2-dz^2=dt^2-dl^2
\end{equation}
have the form\footnote{from now on the summation over repeated
indices $a,b,c,\dots=0,1,2,3$ and $\bar a,\bar b,\bar c=1,2,3$ is
assumed and the indices are lowered and raised in $M_4$ by the
tensors $\eta_{ab}$ and $\eta^{ab}$, respectively} \cite{Shveb}
\begin{equation}\label{Dirsv}
\left(i\gamma^a\overrightarrow\partial_a-m\right)\Psi(x)=0,\qquad
\overline{\Psi}(x)\left(i\overleftarrow\partial_a\gamma^a+m\right)=0,
\end{equation}
where \[\overrightarrow\partial_a\Psi(x)=\partial\Psi/\partial
x^a=\Psi_{,a},\]
\[\overline{\Psi}(x)=\Psi(x)^+\gamma^0,\quad x\in M_4,\] with
$\gamma^a$ ($a=0,1,2,3$) standing for Dirac matrices:
\begin{equation}\label{matDir}
\gamma^0=\left(\matrix{I&0\cr0&-I}\right),\qquad
\gamma^\alpha=\left(\matrix{0&i\sigma^{\bar a}\cr-i\sigma^{\bar
a}&0}\right),
\end{equation}
and $\sigma^{\bar a}$ ($\bar a=1,2,3$) for Pauli matrices:
\begin{equation}\label{matDir}
\sigma^1=\left(\matrix{0&1\cr1&0}\right),\qquad
\sigma^2=\left(\matrix{0&-i\cr i&0}\right),\qquad
\sigma^3=\left(\matrix{1&0\cr0&-1}\right).
\end{equation}

\medskip

The electromagnetic field $A_a(x)$ is introduced as minimal
connection in the covariant derivative
$\nabla_a=\partial_a-ieA_a(x)$ in the equations of motion. Under
these conditions the Dirac equations read:
\begin{equation}\label{Dirvz}
\left(i\gamma^a(\overrightarrow\partial_a-ieA_a)-m\right)\Psi(x)=0,\qquad
\overline{\Psi}(x)\,\left((\overleftarrow\partial_a+ieA_a)i\gamma^a+m\right)=0.
\end{equation}

Let $\gamma(s)$ be a contour in spacetime defined by the equations
$x_a=x_a(s)$ \cite{Fadeev}. The vector field $\dot\gamma(s)$ with
the components $X_a=dx_a/ds$ is tangent to the curve $\gamma(s)$. A
field $\Psi(x)$ is said to be parallel transported along the curve
$\gamma(s)$, if at each its point
$$\nabla_a\Psi(x)X^a|_{\gamma(s)}=0.$$

The variation of the field $\Delta^{12}\Psi(x)$ after parallel
transport along an infinitely small contour $\gamma=\gamma(s)$,
having parallelogram form $(x,x+\Delta_1 x,x+\Delta_1 x+\Delta_2
x,x+\Delta_1x,x)$ is
\begin{equation}\label{izmk}
\Delta_{12}\Psi(x)=F_{ab}\Psi(x)\Delta S_{12}^{ab},
\end{equation}
where $F_{ab}=\partial_bA_a-\partial_aA_b$ stands for the curvature
in the charge space and $\Delta S_{12}^{ab}=\Delta_1x^a\Delta_2x^b-
\Delta_1x^b\Delta_2x^a$ is the area of the
parallelogram outlined by the contour.

\medskip

The field $iA_a(x)$ is a complex function. The duplication of the
algebra of complex numbers leads to a new algebra \cite{octo}. The
most interesting duplication is the algebra of quaternions whose
duplication yields, in particular, the algebra of octonins. In this
case the field $A_a(x)$ can be decomposed over the generators
$\Sigma^{\tilde a},\tilde a=1,2,\dots,7:A_a(x)=A_a^{\tilde
a}(x)\Sigma^{\tilde a}$ having the following properties
\begin{equation}\label{nonas}
\Sigma^{\tilde a}\Sigma^{\tilde b}=\delta^{\tilde a\tilde
b}+i\varepsilon^{\tilde a\tilde b\tilde c}\Sigma^{\tilde c},
\end{equation}
where $\varepsilon^{\tilde a\tilde b\tilde c}$ is completely
antisymmetric symbol, whose only non-vanishing components are
\begin{equation}\label{nonasalg}
\varepsilon^{123}\;=\;\varepsilon^{145}\;=\;\varepsilon^{176}\;=\;\varepsilon^{246}\;=\;
\varepsilon^{257}\;=\;\varepsilon^{347}\;=\;\varepsilon^{365}=1,
\end{equation}
while the values $\Sigma^{\tilde a}$ form a non-commutative
(\ref{nonas}) and non-associative (\ref{asst}) algebra, for which:
\begin{equation}\label{asst}
\{\Sigma^{\tilde a},\Sigma^{\tilde b},\Sigma^{\tilde
c}\}=(\Sigma^{\tilde a}\Sigma^{\tilde b})\Sigma^{\tilde c}-
\Sigma^{\tilde a}(\Sigma^{\tilde b}\Sigma^{\tilde
c})=2\varepsilon^{\tilde a\tilde b\tilde c\tilde d}\Sigma^{\tilde
d},
\end{equation}
taking into account that only the following coefficients of the
completely antisymmetric symbol $\varepsilon^{\tilde a\tilde b\tilde
c\tilde d}$ do not vanish:
\begin{equation}\label{ind}
\varepsilon^{1247}\;=\;\varepsilon^{1265}\;=\;\varepsilon^{2345}
\;=\;\varepsilon^{2376}\;=\;\varepsilon^{3146}\;=\;\varepsilon^{3157}
\;=\;\varepsilon^{4567}\;=1.
\end{equation}

Then the variation of the field $\Delta_{12}\Psi(x)$ under a
parallel transport along the infinitely small contour
$\gamma=\gamma(s)$ also equals (\ref{izmk}), but
$$F_{ab}=\partial_bA_a-\partial_aA_b+i[A_a,A_b].$$

Since the full matrix algebra is associative, the values
$\Sigma^{\tilde a}$ cannot be represented by matrices using
conventional matrix multiplication. Although, following \cite{Zorn},
we may represent them by $4\times4$ matrices
\begin{equation}\label{Zornpr}
\matrix{\Sigma^0=\left(\matrix{1&0\cr0&1}\right),& \Sigma^{\bar
a}=\left(\matrix{0&-i\sigma^{\bar a}\cr i\sigma^{\bar
a}&0}\right),\cr \Sigma^4=\left(\matrix{-1&0\cr0&1}\right),&
\Sigma^{4+\bar a}=\left(\matrix{0&-\sigma^{\bar a}\cr -\sigma^{\bar
a}&0}\right),}
\end{equation}
with the following multiplication law \cite{Daboul}:
\begin{equation}\label{Dab}
\left(\matrix{\lambda&A\cr B&\xi}\right)*
\left(\matrix{\lambda'&A'\cr B'&\xi'}\right)=
\left(\matrix{\lambda\lambda'+\frac12\tr(AB')\hfill&\lambda A'+\xi'
A+\frac i2[B,B']\hfill\cr\lambda'B+\xi B'-\frac
i2[A,A']&\xi\xi'+\frac12\tr(BA')\hfill}\right),
\end{equation}
where $A,A',B,B'$ are real $2\times 2$ matrices and
$\lambda,\lambda',\xi,\xi'$ are scalar real $2\times 2$ matrices.

Introduce the field Lagrangian $A_a(x)$ in the Minkowskian space as
the following scalar:
\begin{equation}\label{lagr}
L=-\frac1{672\pi e_g^2}F_{ab}F^{ab}.
\end{equation}

In the Riemannian space $\Omega_4$ with the metric
$ds^2=g_{\mu\nu}dx^\mu dx^\nu$, a vector $A_\mu(x)$ being parallel
transported along the contour $\gamma(s)$ an infinitely small
distance $dx$ is subject to the variation \cite{Landau}: $\delta
A_\mu=\Gamma^\lambda_{\mu\nu}A_\lambda$, therefore the covariant
derivative of the vector $A_\mu$ in the Riemannian space $\Omega_4$
has the form $\nabla_\nu A_\mu=A_{\mu;\nu}=\partial_\nu
A_\mu+\Gamma^\lambda_{\mu\nu}A_\lambda$. When the vector $A_\mu$ is
parallel transported along the closed contour $\gamma(s)$ in
$\Omega_4$, it changes according to the following formula:
\begin{equation}\label{vvek}
\Delta A_\mu=R^\tau_{\mu\nu\lambda}A_\tau\Delta S^{\nu\lambda},
\end{equation}
where, as above, $\Delta S^{\nu\lambda}$ is an element of the
surface outlined by the contour and $R^\tau_{\mu\nu\lambda}$ is
Riemann tensor:
\begin{equation}\label{tkr}
R^\tau_{\mu\nu\lambda}=\Gamma^\tau_{\mu\lambda,\nu}-
\Gamma^\tau_{\mu\nu,\lambda}+\Gamma^\tau_{\sigma\nu}\Gamma^\sigma_{\mu\lambda}-
\Gamma^\tau_{\sigma}\Gamma^\sigma_{\mu\nu},
\end{equation}

The curvature tensor can also be defined in tetrad representation.
For that, introduce orthogonal basis tetrads $e_\mu^a(x)$ at each
point of $\Omega_4$ defined by the conditions
$$e_\mu^a(x)e_{\nu a}(x)=g_{\mu\nu}(x),\quad
e_a^{\mu}(x)e_{\mu b}(x)=\eta_{ab}(x)$$
and Ricci rotation coefficients
$\gamma_{abc}=e_{a\mu;\nu}e^\mu_be^\nu_c$. Then the tetrad
representation gives the curvature tensor the following form:
\begin{equation}\label{krtet}
R_{abcd}=\gamma_{abc,d}-\gamma_{abd,c}+\gamma_{abf}(\gamma^f_{\ cd}
-\gamma^f_{\ dc})+\gamma_{afc}\gamma^f_{\
bd}-\gamma_{afd}\gamma^f_{\ bc}.
\end{equation}

When a material point causally connected with certain body moves in
Min\-kowskian space, an octonion field $A_a(x)$ gives rise to an
interaction (like in classical electrodynamics), which changes the
initial direction of its motion. In order to evaluate this change,
exact calculations based on octonion algebra should be carried out.
Due to non-associativity of octonion algebra, the direction of the
motion will change as if there is an ``internal curvature" in
Minkowski space which we interpret as Riemannian curvature. In other
words, we assume the gravitation to be induced by the
non-associative part (denote it $\tilde\tr$) as follows:
\begin{equation}\label{main}
-\frac1{672\pi e_g^2}\tilde\tr(F_{ab}F_{cd})=\frac1\kappa
R_{abcd},
\end{equation}
where $R_{abcd}$ is the curvature tensor of the Riemannian space in
tetrad representation. Note that the equation (\ref{main}) is
Lorenz-covariant. It can be verified directly that Bianchi identity
does not hold for the introduced curvature tensor. Therefore, in the
geometrical interpretation of curvature, the Bianchi identity should
be treated as additional dynamical equations when finding
representations of the solution (\ref{main}).

Let us find the non-associative part $\tilde\tr(F_{ab}F_{cd})$ in
case of classical fields. Since the associator is calculated on at
least three fields, we exclude the terms proportional to the square
of $A_a$. Since the matrices $\Sigma^{\tilde a}$ have zero trace,
the associator of three matrices $\Sigma^{\tilde a}$ vanishes, as
(\ref{main}) can contain at least four matrices $\Sigma^{\tilde a}$.
In the meantime
\begin{equation}\label{ass4}
\tilde\tr\{\Sigma^{\tilde a},\Sigma^{\tilde b},\Sigma^{\tilde
c},\Sigma^{\tilde d}\}=\tr(\{\Sigma^{\tilde a},\Sigma^{\tilde
b},\Sigma^{\tilde c}\}\Sigma^{\tilde d}-\Sigma^{\tilde a}\{\Sigma^{\tilde b},
\Sigma^{\tilde c},\Sigma^{\tilde d}\})=8\varepsilon^{\tilde
a\tilde b\tilde c\tilde d}.
\end{equation}

\noindent Therefore,
\begin{equation}\label{maincl}
\frac1\kappa R_{abcd}=-\frac1{168\pi e_g^2} \varepsilon^{\tilde
a\tilde b\tilde c\tilde d}\left(A_a^{\tilde a}A_b^{\tilde
 b}-A_b^{\tilde a}A_a^{\tilde b}\right) \left(A_c^{\tilde c}A_d^{\tilde d}-
 A_d^{\tilde c}A_c^{\tilde d}\right).
\end{equation}

\section{Uniform isotropic space}

Consider a uniform isotropic space with the Friedmann metric
\begin{equation}\label{frid}
ds^2=a^2(\eta)(d\eta^2-dl^2)
\end{equation}
for which the only non-vanishing components of the curvature tensor
are
\begin{equation}\label{kr}
R_{0\alpha0\beta}=\frac{a'^{\,2}-a''a}{a^4}\eta_{\alpha\beta},\qquad
R_{\alpha\beta\gamma\delta}=\frac{a'^{\,2}}{a^4}
(\eta_{\beta\gamma}\eta_{\alpha\delta}-
\eta_{\alpha\gamma}\eta_{\beta\delta})
\end{equation}
Substitute (\ref{kr}) into (\ref{maincl}) and denote
$\theta^4=\kappa/168\pi e_g^2$, then
$$\frac{a'^{\,2}-a''a}{a^4\theta^4}\eta_{\alpha\beta}=
-\varepsilon^{\tilde a\tilde b\tilde c\tilde d}\left(A_0^{(\tilde
a)}A_\alpha^{(\tilde b)}-A_\alpha^{(\tilde a)}A_0^{(\tilde
b)}\right) \left(A_0^{(\tilde c)}A_\beta^{(\tilde
d)}-A_\beta^{(\tilde c)}A_0^{(\tilde d)}\right)=$$
$$=\left(A_0^{(1)}A_\alpha^{(2)}-A_\alpha^{(1)}A_0^{(2)}\right)
\left(A_0^{(4)}A_\beta^{(7)}-A_\beta^{(4)}A_0^{(7)}\right)+$$
\begin{equation}\label{krkl}
+\left(A_0^{(4)}A_\alpha^{(7)}-A_\alpha^{(4)}A_0^{(7)}\right)
\left(A_0^{(1)}A_\beta^{(2)}-A_\beta^{(1)}A_0^{(2)}\right)+\dots
\end{equation}
$$\frac{a'^{\,2}}{a^4\theta^4}(\eta_{\beta\gamma}\eta_{\alpha\delta}-
\eta_{\alpha\gamma}\eta_{\beta\delta})= -\varepsilon^{\tilde a\tilde
b\tilde c\tilde d}\left(A_\alpha^{(\tilde a)}A_\beta^{(\tilde
b)}-A_\beta^{(\tilde a)}A_\alpha^{(\tilde b)}\right)
\left(A_\gamma^{(\tilde c)}A_\delta^{(\tilde d)}-A_\delta^{(\tilde
c)}A_\gamma^{(\tilde d)}\right)=$$
$$=\left(A_\alpha^{(1)}A_\beta^{(2)}-A_\beta^{(1)}A_\alpha^{(2)}\right)
\left(A_\gamma^{(4)}A_\delta^{(7)}-A_\delta^{(4)}A_\gamma^{(7)}\right)+$$
\begin{equation}\label{krkl2}
+\left(A_0^{(4)}A_\alpha^{(7)}-A_\alpha^{(4)}A_0^{(7)}\right)
\left(A_0^{(1)}A_\beta^{(2)}-A_\beta^{(1)}A_0^{(2)}\right)+\dots,
\end{equation}
where the dots stand for summation over all indices (\ref{ind}).
Denote \[x^2=(a'^2-a''a)/a^4\theta^4,\qquad y^4=a'^2/a^4\theta^4\]
and
\[A_\alpha^{(\tilde
a)}=\tilde A_\alpha^{(\tilde a)}y,\qquad A_0^{(\tilde a)}=\tilde
A_0^{(\tilde a)}x/y,\] then we obtain the system
$$\left\{\matrix{\varepsilon^{\tilde
a\tilde b\tilde c\tilde d}\left(\tilde A_0^{(\tilde a)}\tilde
A_\alpha^{(\tilde b)}-\tilde A_\alpha^{(\tilde a)}\tilde
A_0^{(\tilde b)}\right) \left(\tilde A_0^{(\tilde c)}\tilde
A_\beta^{(\tilde d)}-\tilde A_\beta^{(\tilde c)}\tilde A_0^{(\tilde
d)}\right)= -\eta_{\alpha\beta}\hfill\cr \varepsilon^{\tilde a\tilde
b\tilde c\tilde d} \left(\tilde A_\alpha^{(\tilde a)}\tilde
A_\beta^{(\tilde b)}- \tilde A_\beta^{(\tilde a)}\tilde
A_\alpha^{(\tilde b)}\right) \left(\tilde A_\gamma^{(\tilde
c)}\tilde A_\delta^{(\tilde d)}-\tilde A_\delta^{(\tilde c)}\tilde
A_\gamma^{(\tilde d)}\right)=
-\eta_{\beta\gamma}\eta_{\alpha\delta}+
\eta_{\alpha\gamma}\eta_{\beta\delta},}\right.$$ whose solutions
$A^{(a)}_i$ have the form:
$$A^{(\tilde
a)}_0=\frac1{a\theta}\sqrt{a'-a''a/a'}\cdot\Co^{\tilde{a}}_0,\qquad
A^{(\tilde a)}_\alpha=\frac1{a\theta}\sqrt{a'}\cdot\Co^{\tilde
a}_\alpha.$$

For the state of the matter $p=\varepsilon/3$ when
$a(\eta)=C\cdot\eta$, we get the so\-lu\-tions:
$$A^{(\tilde a)}_k=\Co^{\tilde a}_k/\eta.$$

In case of dust-like matter $p=0$ when $a(\eta)=C\cdot\eta^2$ we
obtain: $$A^{(\tilde a)}_k=\Co^{\tilde a}_k\cdot\eta^{-3/2}.$$

\section{Concluding remarks}

We have assumed that in classical case matter fields induce the
field of octonions, which, in turn, interacts with the matter in
such a way that in order to correctly describe the motion in terms
of associative algebra a Riemannian metric needs to be introduced.
As a result, from a formal perspective, there is no need to
recalculate the known effects of General Relativity which might be
different in the suggested model. However, the form of the equation
(\ref{maincl}) may restrict the possible geometry of physical space.

The author expresses his gratitude to the participants of the
Friedmann Seminar for Theoretical Physics (St. Petersburg) for
profound discussions. The work was carried with the support of the
Russian Ministry of Education (grant RNP No 2.1.1.68.26).


\begin{thebibliography}{99}

\bibitem{Baez}
J.~C.~Baez, \emph{The octonions}, Bull. Amer. Math. Soc., {\bf 39},
145--205 (2002), math.ra/0105155.

\bibitem{Zorn}
Max~Zorn, \emph{Alternativkorper und quadratische Systeme}, Abh.
Math. Sem. Univ. Hamburg, {\bf 9}, 395--402 (1933).

\bibitem{Daboul}
J.~Daboul, R.~Delbourgo, \emph{Matrix Representation of Octonions
and Generalizations}, J. Math. Phys., \textbf{40} 4134--4150 (1999),
hep-th/9906065.

\bibitem{Shveb}
S.~Schweber, \emph{An Introduction to Relativistic Quantum Field
Theory}, Harper and Row, New York (1961).

\bibitem{Fadeev}
L.~D.~Faddeev and A.~A.~Slavnov, \emph{Gauge fields, Introduction to
Quantum Theory}, Benjamin/Cummings, New York (1980).

\bibitem{octo}
T.~A.~Springer, F.~D.~Veldkamp. \emph{Octonions, Jordan algebras and
exceptional groups}. Springer-Verlag, Berlin (2000).

\bibitem{Landau}
L.~D.~Landau and E.~M.~Lifschitz, \emph{The Classical Theory of
Fields}, Butterworth-Heinemanm, Oxford, UK (1987).

\end{thebibliography}
\end{document}